\documentclass[%
 preprint, 
 amsmath,amssymb,
 aps,
pra,
superscriptaddress,
]{revtex4-2}
\usepackage{graphicx}
\usepackage{dcolumn}
\usepackage{bm}
\usepackage{float}
\usepackage{amsfonts}
\usepackage{color}
\usepackage[english]{babel}
\usepackage[usenames,dvipsnames]{xcolor}
\usepackage{array}
\usepackage[pagewise]{lineno}
\begin{document}

\title{Classical-trajectory Monte Carlo calculations of differential electron emission in fast heavy-ion collisions with water molecules}

\author{Alba Jorge}
\email{albama@yorku.ca}
\author{Marko Horbatsch}
\affiliation{Department of Physics and Astronomy, York University, Toronto, Ontario, M3J 1P3, Canada}
\author{Clara Illescas}
\affiliation{Departamento de Qu\'imica, Universidad Aut\'onoma de Madrid, Madrid, 28049, Spain.}%

\author{Tom Kirchner}
\affiliation{Department of Physics and Astronomy, York University, Toronto, Ontario, M3J 1P3, Canada}




\begin{abstract}
A classical description of electron emission differential ionization cross sections for highly-charged high-velocity ions ($\sim$ 10 a.u.) impinging on water molecules is presented. We investigate the validity of the classical statistical mechanics description of ionization ($\hbar=0$ limit of quantum mechanics) in different ranges of electron emission energy and solid angle, where mechanisms such as soft and binary collisions are expected to contribute. The classical-trajectory Monte Carlo method is employed to calculate doubly and singly differential cross sections for C$^{6+}$, O$^{8+}$ and Si$^{13+}$ projectiles, and comparisons with Continuum Distorted Wave Eikonal Initial State theoretical results and with experimental data are presented. We implement a time-dependent screening effect in our model, in the spirit of mean-field theory to investigate its effect for highly charged projectiles. We also focus on the role of an accurate description of the molecular target by means of a three-center potential to show its effect on differential cross sections. Very good agreement with experiments is found at medium to high electron emission energies.
\end{abstract}
\maketitle
\section{Introduction}
The investigation of inelastic processes involving biological molecules under the impact of highly charged ions is an active field of study, mainly due to its relevance for hadron therapy \cite{w46,ak05}.
A major research objective is the full step-by-step description of the particles involved and their dynamics following the irradiation of matter. Electrons ejected from water molecules due to direct ionization by the projectile or due to the Auger effect, will cause secondary interactions with the biological medium. Depending on the electron energies they will lead to further ionization or excitation processes \cite{sde10}. The microscopic understanding of the whole time evolution of the heavy-ion impact is needed to accurately describe the final energy deposition, which leads to damage to the DNA of the irradiated tumor cell \cite{bch+00}. Monte Carlo codes are used to simulate all these interactions using the differential and total cross sections for the inelastic processes of different projectiles impinging on water molecules \cite{cbc08}. 

A large number of studies can be found for theoretical and experimental investigations of total ionization cross sections as well as the subsequent fragmentation of the water molecule due to ion impact \cite{rgd+85,wbb+95,gec+04,ahs05,sph+06,bl07,iem+11,mkh+12a,mkh+12b,lwm+16}. Fewer studies have been devoted to differential cross sections, and, in general, these have focused on low-charge projectiles at intermediate energies \cite{tw77,twp80,br86,oso+05,okh+05f}. However, in recent years, more attention has been given to highly charged projectiles and more energetic collisions \cite{cbc+07,dcb+09,nbk+13,tmf+14,bbb+16,bbm+17}.
Although most of the recent studies have been devoted to the water molecule, the effects of charged particles colliding with other molecules of biological interest are being investigated as well \cite{mp06,tef+10b,ikii11,ank+13,khj+16}.

In this work we look at the validity of a classical description of water molecule ionization, more specifically in the area of angular- and energy-dependent ionized electron distributions.
Previously, the classical-trajectory Monte Carlo (CTMC) method \cite{ap66} has been applied to molecules through the combination with the over-barrier method \cite{acz+08,lac+09}, with single-center potentials \cite{eim+07}, or in a more sophisticated way with the use of multi-center potentials \cite{gei+10b,iem+11,khj+16,s15}.
A three-center potential \cite{gem+12,eim+15} was demonstrated to describe the interaction of the frozen H$_2$O$^+$ core with the active electron accurately. When combined with the CTMC method, it provides total cross sections which compare quite well with experimental values for H$^+$, He$^{2+}$ and C$^{6+}$ projectiles, in the collision energy range from 10 keV/u to 5 MeV/u \cite{iem+11,eim+13}. In order to determine the validity of this method when dealing with the differential description of ionization, we compare in this work new results with recent experimental and theoretical data from \cite{bbb+16,bbm+17,bbc+18}, for C$^{6+}$, O$^{8+}$ and Si$^{13+}$ projectiles at impact energies between 3 and 4 MeV/u. Special emphasis is put on the study of the so-called saturation effect, when for high $Z_p/v_p$ (projectile charge to impact velocity ratio, in a.u.) the measured electron emission increases at a lesser rate than predicted by the Bethe-Born model \cite{bbc+18}. We study this effect for systems in the range $0.47\leq Z_p/v_p\leq 1.03$ and, in order to do so, we have introduced a time-dependent charge-state target potential in our calculations. We analyze the net results, for both doubly and singly differential electron emission (with respect to ionized electron energy and angle), and the contributions associated with a fixed number of continuum electrons (multiple ionization contributions). We also report on an analysis of the contributions coming from the different molecular orbitals, which have different ionization energies and spatial and momentum distributions of their electronic clouds.

Another aspect we focus on is the effect of the multi-center nature of the water molecule on the DDCS as a function of the ejection angle. In our CTMC model the target is represented by a three-center model potential which accounts for both the passive electrons and also for the three nuclear charges in the water molecule. A direct comparison with the post and prior continuum distorted wave eikonal initial state (CDW-EIS) results of Refs. \cite{bbb+16,bbm+17,bbc+18} is particularly interesting to us. In the CDW-EIS model for ion-atom collisions the active electron is described by a two-center wavefunction, given by the product of a target bound orbital (at the Hartree-Fock level) and a projectile continuum eikonal phase in the entry channel, and by a product of a plane wave and two factors associated with Coulombic target and projectile fields in the exit channel \cite{bbc+18}. Post and prior versions differ in that only the latter takes the residual interaction of the active (continuum) and the passive (bound) electrons into account. This interaction is referred to as dynamical screening in Refs. \cite{bbb+16,bbm+17,bbc+18}.
The molecular character of the H$_2$O target system is modeled using the 'complete neglect of differential overlap' (CNDO) approach in which the molecular cross section is written as a linear combination of atomic contributions with weight factors that are obtained from a Mulliken-like population analysis. We can extract information about multi-center target effects by comparing results from these CDW-EIS CNDO calculations with our CTMC model, especially by looking at the forward-backward asymmetry in the DDCS as a function of the electron ejection angle.

The layout of the paper is as follows: first, a brief explanation of the theory is presented, which comprises the special features in the CTMC model for ion-molecule systems (section \ref{ctmc}), followed by the calculation of differential probabilities (\ref{diff}) and then the time-dependent screening approach (\ref{dyn}). In section \ref{qfold} we present expressions used to analyze $q$-fold ionization of the molecule ($q=1,2,3$) with one electron observed in emission angle and energy. The obtained angular and energy distributions are compared to experimental and theoretical data \cite{bbb+16,bbm+17,bbc+18} in the Results section (\ref{results}), and finally we give summarizing comments and conclusions in section (\ref{conclusiones}).

Atomic units ($\hbar=m_e=q_e=4\pi\epsilon_0=1$) are used throughout unless otherwise stated.
\section{Theoretical method\label{ctmc}}
The CTMC method as applied to ion-H$_2$O collisions was described recently in Refs. \cite{iem+11,khj+16}. This variant of the method makes use of the semi-classical impact parameter approximation (which is certainly valid for high-energy ion-molecule collisions) and then approximates the electron dynamics with a mean-field potential model. It can be understood as a replacement of the Schr\"odinger description by an $\hbar=0$ limit, i.e. quantum mechanics for the electron motion is approximated by classical statistical mechanics.
The initial distributions for the molecular orbitals (MO) are generated as described in \cite{iem+11} and are composed of trajectories bound to a three-center potential, which describes H$_2$O in a mean-field sense, with fixed values of the ionization energies ($E_{1b_1}=$ 0.5187 a.u., $E_{3a_1}=$ 0.5772 a.u., $E_{1b_2}=$ 0.7363 a.u., $E_{2a_1}=$ 1.194 a.u., $E_{1a_1}=$ 20.25 a.u.). This microcanonical initial distribution contains 5$\cdot 10^5$ trajectories per MO. The time evolution for the five electrons is performed simultaneously so that a time-dependent net ionization probability can be calculated at each time step for the entire ensemble, which is needed for the time-dependent screening approach (see section \ref{dyn}). The electron model potential is given by adding three spherical contributions \cite{rem+08}:
\begin{equation} V_{\rm{mod}}=V_{\rm{O}}(r_{\rm{O}})+V_{\rm{H}}(r_{\rm{H_1}})+V_{\rm{H}}(r_{\rm{H_2}}).\end{equation}
\begin{equation}
\begin{alignedat}{1}
&V_{\rm{O}}(r_{\rm{O}})=-\frac{8-N_{\rm{O}}}{r_{\rm{O}}}-\frac{N_{\rm{O}}}{r_{\rm{O}}}(1+\alpha_{\rm{O}}r_{\rm{O}})\exp(-2\alpha_{\rm{O}}r_{\rm{O}})\\
&V_{\rm{H}}(r_{\rm{H_i}})=-\frac{1-N_{\rm{H}}}{r_{\rm{H}}}-\frac{N_{\rm{H}}}{r_{\rm{H}}}(1+\alpha_{\rm{H}}r_{\rm{H}})\exp(-2\alpha_{\rm{H}}r_{\rm{H}})\\
\end{alignedat}.
\label{pot}
\end{equation}
where $N_{\rm{O}}=7.185$, $\alpha_{\rm{O}}=1.602$, $N_{\rm{H}}=(9-N_O)/2$, $\alpha_{\rm{H}}=0.6170$. Here, $r_{\rm{O}}$, $r_{\rm{H}}$ are defined as the electron distances from the respective nuclei. The projectile potentials are all represented using a Coulomb form $V_p=-\frac{Z_p}{r_p}$, with $r_p$ the electron distance to the projectile, even for Si$^{13+}$, which is treated as a bare projectile. 

The time evolution of the initial phase-space distributions is given by Hamilton's equations. The collision dynamics caused by the projectile is represented by a potential function following a rectilinear trajectory. The integration in time can in general be stopped at around 500 a.u. of distance between the target and projectile, although for some trajectories this is not sufficient, since the less energetic ionized electrons take more time to become insensitive to the separating charges, namely the projectile and the residual water molecule ion. For these trajectories integration in time has been carried out until these interactions can be neglected ($|E_{\rm{el}}(t)-E_{\rm{el}}(t+\Delta t)|/E_{\rm{el}}(t)< 10^{-4}$), with $E_{\rm{el}}=p^2/2+V_{\rm{mod}}+V_p$. The orientation of the molecule is changed for each initial trajectory by rotating the molecule using random Euler angles.
\subsection{Emission energy and angle differential probabilities\label{diff}} 
In order to determine the doubly and singly differential cross sections, DDCS and SDCS respectively, we define 'boxes', for both $\Delta E_{\rm el}$ and $\Delta \Omega_{\rm el}$, in which the ionized electrons are binned. The probability associated with given emission angle and energy is then \cite{khj+16}
\begin{equation} p=\frac{n}{N_T \Delta E_{\rm el} \Delta \Omega_{\rm el}}\:\:.\end{equation}
where $\Delta \Omega_{\rm el}=\sin(\theta_{\rm el})\Delta \theta_{\rm el} \Delta \phi_{\rm el}$ and $(\theta_{\rm el},\phi_{\rm el})$ are the polar and azimuthal angles of the emitted electron's momentum vector with respect to the initial projectile momentum, $n$ is the number of electrons emitted at the given $E_{\rm{el}}$ and $\Omega_{\rm el}$ and $N_T$ is the total number of initial trajectories. Due to the symmetry with respect to the azimuthal angle, $\Delta \Omega_{\rm el}=2\pi[\cos(\theta_{\rm el_i})-\cos(\theta_{\rm el_{i+1}})]$, where $\theta_{\rm el_i}$ and $\theta_{\rm el_{i+1}}$ define the interval for a given $\theta_{\rm el}$: $\theta_{\rm el_i}\leq \theta_{\rm el}< \theta_{\rm el_{i+1}}$. The total probability is then
\begin{equation} \frac{{\rm{d}}^2P}{{\rm{d}} E_{\rm{el}}{\rm{d}}\Omega_{\rm{el}}}=2\sum_{j=1}^{5}\frac{n_j}{N_{T_j} \Delta E_{\rm el} 2\pi[\cos(\theta_{\rm el_i})-\cos(\theta_{\rm el_{i+1}})]}\:\:.\end{equation}
The index $j$ enumerates the MOs and the factor of two accounts for the fact that each MO is occupied by two electrons due to spin degeneracy. In the same manner, the probabilities for computing the SDCS in energy or angle can be obtained
\begin{equation} 
\begin{alignedat}{1}
&\frac{{\rm{d}}P}{{\rm{d}} E_{\rm{el}}}=2\sum_{j=1}^{5}\frac{n_j}{N_{T_j} \Delta E_{\rm el}}\:\:,\\
&\frac{{\rm{d}}P}{{\rm{d}}\Omega_{\rm{el}}}=2\sum_{j=1}^{5}\frac{n_j}{N_{T_j} 2\pi[\cos(\theta_{\rm el_i})-\cos(\theta_{\rm el_{i+1}})]}\:\:.
\end{alignedat}
\end{equation}
The sizes of the boxes $\Delta E_{\rm el}$ and $\Delta \Omega_{\rm el}$ depend on the cross section we look at. For example, when dealing with the DDCS for a low-emission energy, like 5 eV, $\Delta E_{\rm el}=1$ eV, so in this case $E_{\rm{el}}=5$ $\pm 0.5$ eV. In the case of a high-emission energy, like 200 eV, we use $\Delta E_{\rm el}=10$ eV (and consequently, $E_{\rm el}=200\pm 5$ eV). When dealing with DDCS as a function of the emission energy, the $\Delta E$ depends on $E$ logarithmically, with $\Delta E=1$ eV for $E_{\rm el}<10$ eV, $\Delta E=10$ eV for $10<E_{\rm el}<100$ eV and $\Delta E=100$ eV for $100<E_{\rm el}<1000$ eV. With respect to the $\Delta \Omega_{\rm el}$ we have used values of 6 to 20 degrees, depending on the forward-backward asymmetry and the statistics. For example, at 5 eV a bin size of 20$^\circ$ is sufficient while for 200 eV we reduce it to $\Delta \theta_{\rm el}=10^\circ$. With respect to the DDCS as a function of the emission energy for a given emission angle, we have used $\Delta \theta_{\rm el}=6^\circ$ so, for example, we have $\theta_{\rm el}=90\pm 3^\circ$.
For each impact parameter, final probabilities for ionization and electron capture are computed at the end of the collision ($t\geq 500/v$ a.u.) by looking at the final energies of the active electron with respect to both target and projectile. Probabilities for $q$-fold ionization are obtained within the Independent Electron Model (IEM), computed from a binomial analysis of these single probabilities.
\subsection{Time-dependent screening\label{dyn}}
We also investigate the saturation behavior of the net ionization cross section in collisional systems for which the projectile charge to velocity ratio $Z_p/v_p\gtrsim 1$ \cite{bbc+18}. In the IEM the multiple ionization probablities are obtained through a binomial analysis based on the single-ionization probability, and can present limitations when dealing with multiple-electron transitions. The amount of ionization during the dynamics grows for highly-charged projectiles, causing non-negligible values of $q$-fold, $q>1$, ionization of the water molecule, which is questioning the validity of the static frozen-electron potential. 

In order to investigate this problem, we propose a time-dependent screening approach \cite{khl+00} (called dynamical screening in that work) for the three-center potential, based on decreasing the local screening parameters at each center according to the level of ionization during the collision. The $N_O$ and $N_H$ values in the three-center potential correspond to the expected average number of frozen electrons, 9 in total, at each center, $\sim$7.2 and $\sim$0.9 for the oxygen and hydrogen atoms, respectively. The screening charge parameters add up to nine in order to yield the correct asymptotic behavior of the effective potential. A simple model for the time-dependent mean field screening is achieved by making these $N_i$ parameters time-dependent, so that the amount of frozen electrons at each center decreases if the ionization probability $p_j(t)$ increases to values where $q$-fold, $q>1$, ionization is non-negligible. Therefore, $N_i=N_i[p_j(t)]$ and the constant values $N_{\rm{O}}$ and $N_{\rm{H}}$ from Eq. (\ref{pot}) are renamed as $N_{\rm{O}}^c$ and $N_{\rm{H}}^c$. The application of Eq. (14) from \cite{khl+00} to the three-center potential implies a decrease of $N_i[P_{\rm{Net}}(t)]$, with $P_{\rm{Net}}(t)=\sum\limits_{j=1}^{5}2p_j$, that is linear after the $q=1$ threshold, which can be described by:
\begin{equation}
N_{\rm{O}}({\rm{P}}_{\rm{Net}})=
\begin{cases}
	\begin{alignedat}{2}
		&N_{\rm{O}}^c\:\:\:\:\:\:\:\:\:\:&&{\rm{P}}_{\rm{Net}}\leq 1\\
		&a\cdot 8(1-0.1{\rm{P}}_{\rm{Net}})\:\:\:\:\:\:\:\:\:\:&& 1<{\rm{P}}_{\rm{Net}}\leq 10\\
	\end{alignedat}
\end{cases}
\end{equation}

\begin{equation}
N_{\rm{H}}({\rm{P}}_{\rm{Net}})=
\begin{cases}
	\begin{alignedat}{2}
		&N_{\rm{H}}^c\:\:\:\:\:\:\:\:\:\:&&{\rm{P}}_{\rm{Net}}\leq 1\\
		&b\cdot (1-0.1{\rm{P}}_{\rm{Net}})\:\:\:\:\:\:\:\:\:\:&& 1<{\rm{P}}_{\rm{Net}}\leq 10\\
	\end{alignedat}
\end{cases}
\end{equation}
with the $a=7.185/7.2$, $b=0.9075/0.9$ factors used to assure the continuity of the $N_i({\rm{P}}_{\rm{Net}})$ functions. The net ionization probability is calculated in each time step, using an initial distribution which considers all the MOs simultaneously and for which the dynamics is evolved concurrently. The $N_i[{\rm{P}}_{\rm{Net}}(t)]$ remain constant and equal to the initial values for ${\rm{P}}_{\rm{Net}}(t)\leq 1$, and decrease to zero for ${\rm{P}}_{\rm{Net}}(t)=10$, when the potential reduces to a pure three-center Coulomb potential.
\subsection{$q$-fold ionization\label{qfold}}
In order to analyze the influence of the time-dependent screening in the studied systems, we investigate the importance of the separate $q$-fold ionization contributions for given impact parameter $b$. The total probabilities for the release of $q$ electrons are given by \cite{kgm+02}
\begin{equation} P_q(b)=\sum_{\substack{q_1,...,q_m=0 \\ q_1+...q_m=q}}^{N_1,...,N_m} \prod_{i=1}^{m}\frac{N_i!}{q_i!(N_i-q_i)!}[p_i(b)]^{q_i}[1-p_i(b)]^{N_i-q_i}.\label{pq}\end{equation}
Here $m$ is the number of shells (MOs) and $N_i$ is the number of electrons in each shell, which in the present case equals 2 for each of the five MOs.

In the case of the differential probability of detection of one electron at a given \{$\Delta E_{\rm{el}},\Delta \Omega_{\rm{el}}$\} when ionizing $q$ electrons, we show the $q=$1, 2 and 3 cases, which can be generalized for higher ionization states. In the case of $q=1$
\begin{equation}
\frac{{\rm{d}}^2P_{q=1}^{\rm (s)}}{{\rm{d}} E_{\rm{el}}{\rm{d}}\Omega_{\rm{el}}} =  2 \sum_{ j =1}^m{\frac{{\rm{d}}^2 p_{j}}{{\rm{d}} E_{\rm{el}}{\rm{d}}\Omega_{\rm{el}}} (1-p_{j})
\prod_{k \ne j}^m {(1-p_k)^2}}.\label{pdq1}
\end{equation}
As before, the $p_j$ are the single-particle ionization probabilities for the molecular orbitals (shells). Equation (\ref{pdq1}) is based on the binomial probability of single (one-fold) ionization from the $j^{\rm th}$ MO, i.e., $2 p_j (1-p_j)$ to be multiplied by non-ionization from all other MOs. 

For the case of $q=2$ we have two contributions, namely two electrons ionized from separate MOs (labeled ss), and a double ionization from the same MO (labeled d).
The former can be expressed as
\begin{equation}
\frac{{\rm{d}}^2P_{q=2}^{\rm (ss)}}{{\rm{d}} E_{\rm{el}}{\rm{d}}\Omega_{\rm{el}}} = 2 \sum_{\{ j_1 \ne j_2\} \in \{1..m\}}{\frac{{\rm{d}}^2 p_{j_1}}{{\rm{d}} E_{\rm{el}}{\rm{d}}\Omega_{\rm{el}}} p_{j_2}(1- p_{j_1})(1- p_{j_2})
\prod_{k \notin \{ j_1, j_2\}}^m {(1-p_k)^2}}.
\label{pdq2-1}
\end{equation}
This is the product of two binomial probabilities averaged over both distinct MOs participating in the ionization process (thus the factor of 4 is reduced to 2) times the non-ionization probability for the other MOs.

The double ionization process from a single shell with one of the two electrons analyzed in emission energy and solid angle is straightforward:
\begin{equation}
\frac{{\rm{d}}^2P_{q=2}^{\rm (d)}}{{\rm{d}} E_{\rm{el}}{\rm{d}}\Omega_{\rm{el}}} =  \sum_{ j_1 =1}^m{\frac{{\rm{d}}^2 p_{j_1}}{{\rm{d}} E_{\rm{el}}{\rm{d}}\Omega_{\rm{el}}} p_{j_1}
\prod_{k \ne j_1}^m {(1-p_k)^2}}.\label{pdq2-2}
\end{equation}
One has to add both contributions and obtains
\begin{equation}
\frac{{\rm{d}}^2P_{q=2}}{{\rm{d}} E_{\rm{el}}{\rm{d}}\Omega_{\rm{el}}} = \frac{{\rm{d}}^2P_{q=2}^{\rm (ss)}}{{\rm{d}} E_{\rm{el}}{\rm{d}}\Omega_{\rm{el}}} +\frac{{\rm{d}}^2P_{q=2}^{\rm (d)}}{{\rm{d}} E_{\rm{el}}{\rm{d}}\Omega_{\rm{el}}}.\label{pdq2}
\end{equation}

In the case of $q=3$ electrons ejected from separate shells $j_1, j_2, j_3$, and one of them analyzed in emission solid angle $\Omega_{\rm{el}}$ and energy $E_{\rm{el}}$, an averaging process leads to the following answer when using the binomial expressions for single ionization from each of the three shells (labeled sss):

\begin{equation} 
\frac{{\rm{d}}^2P_{q=3}^{\rm (sss)}}{{\rm{d}} E_{\rm{el}}{\rm{d}}\Omega_{\rm{el}}} = \frac{8}{3} \sum_{\{ j_1 \ne j_2 \ne j_3\} \in \{1..m\}}{\frac{{\rm{d}}^2 p_{j_1}}{{\rm{d}} E_{\rm{el}}{\rm{d}}\Omega_{\rm{el}}} p_{j_2} p_{j_3}(1- p_{j_1})(1- p_{j_2})(1- p_{j_3})
\prod_{k \notin \{ j_1, j_2, j_3\}}^m {(1-p_k)^2}}.\label{pdq3-1}
\end{equation}

The interpretation of this expression is that one takes the product of three binomial single-ionization expressions $2 p_j (1-p_j)$, differentiates one of the three ionization
probabilities with respect to solid angle and energy, and then averages (hence the factor $1/3$), so that the analyzed electron is from any of the three shells considered. This is multiplied by the non-ionization probability of the inactive shells.

For the process where two electrons are ionized from one MO (shell), while the third one is from a different MO, which also contributes at $q=3$ (with label ds), we need to average the differential analysis over the two contributors resulting in a factor of $1/2$. Thus, the binomial ionization probability 
$2 p_1(1-p_1) p_2^2 = (p_1 p_2^2+ p_2 p_2 p_1 )(1-p_1)$ is differentiated in the first of the triple products to yield the general formula for
the net ionization probability with one electron analyzed, and a sum over all MOs (shells) is executed:
\begin{equation} 
\frac{{\rm{d}}^2P_{q=3}^{\rm (ds)}}{{\rm{d}} E_{\rm{el}}{\rm{d}}\Omega_{\rm{el}}} =  \sum_{\{ j_1 \ne j_2 \} \in \{1..m\}}{\frac{{\rm{d}}^2 p_{j_1}}{{\rm{d}} E_{\rm{el}}{\rm{d}}\Omega_{\rm{el}}} \left [ p_{j_2}^2 (1- p_{j_1})+p_{j_1} p_{j_2} (1- p_{j_2}) \right ]
\prod_{k \notin \{ j_1, j_2\}}^m {(1-p_k)^2}}.\label{pdq3-2}
\end{equation}
For the overall $q=3$ result we need to add, i.e.,
\begin{equation} 
\frac{{\rm{d}}^2P_{q=3}}{{\rm{d}} E_{\rm{el}}{\rm{d}}\Omega_{\rm{el}}} = \frac{{\rm{d}}^2P_{q=3}^{\rm (sss)}}{{\rm{d}} E_{\rm{el}}{\rm{d}}\Omega_{\rm{el}}} +\frac{{\rm{d}}^2P_{q=3}^{\rm (ds)}}{{\rm{d}} E_{\rm{el}}{\rm{d}}\Omega_{\rm{el}}}.\label{pdq3}\end{equation}
Integration over emission energy and angles of Eqs. (\ref{pdq1}), (\ref{pdq2}) and (\ref{pdq3}) yields Eq. (\ref{pq}) for $q=$1, 2 and 3. For the rest of $q$-fold equations the same process has to be followed, considering for each term the number of orbitals involved in order to average correctly. As can be seen in equations (\ref{pdq1}), (\ref{pdq2-1}), (\ref{pdq3-1}) and (\ref{pdq3-2}) this implies using an average factor of one divided by the number of shells involved in each term.
\section{Results\label{results}}
We present results for the C$^{6+}$ at 4 MeV/u ($Z_p/v_p=0.47$), O$^{8+}$ at 3.75 MeV/u ($Z_p/v_p=0.65$), O$^{8+}$ at 3 MeV/u ($Z_p/v_p=0.73$) and Si$^{13+}$ at 4 MeV/u ($Z_p/v_p=1.03$) systems. 
\subsection{Angular distributions}
We begin with the DDCS for fixed electron energy as a function of the electron emission angle. This allows us to understand the different ionization mechanisms which contribute to different emission energies and angles. We compare first in Figs. \ref{ddcs200} and \ref{ddcs5} the DDCS as a function of electron ejection angle $\theta_{\rm el}$ for two electron energies, $E_{\rm{el}}=$ 200 and 5 eV. These two energies are representative of the main ionization mechanisms, namely the Soft Collision (SC), Two-center Electron Emission (TCEE) and Binary Encounter (BE) processes. 

Soft electrons, i.e. the less energetic ones, are characterized by being ejected in collisions with large impact parameters. They involve small momentum transfer, and the de Broglie wave length of the SC electrons is comparable with the dimension of the atom: thus a quantum-mechanical treatment is needed. Another general feature of SC electrons is that they are initially ejected predominantly near 90$^\circ$, then they get deflected in the field of the target nucleus and eventually a nearly isotropic emission is found. 

The BE electrons are emitted in small-impact-parameter binary collisions, and thus the process is dominated by the direct interaction of the electrons with the projectile. The TCEE is the ionization mechanism giving rise to a strong forward-backward asymmetry in the DDCS as a function of the ejection angle. Its main characteristic is the influence of the two collision partners, target and projectile, on the dynamics of an electron transferred to the continuum. Due to the attraction of the ionized electrons to the projectile the DDCS are higher at forward angles than in the backward direction (cf. Chapter 2 in Ref. \cite{sdr97}).

In the DDCS as a function of the emission angle the main differences due to the mechanisms described above can be found. We present CTMC results for which the target has been treated with a three-center potential, and consequently we can look not only at the influence of both target and projectile but also at the effect that a multi-center target can have on the results. We propose to extend the TCEE terminology from ion-atom collisions to the Many-Center Electron Emission (MCEE) mechanism, appropriate for molecules. We also keep in mind that CTMC cannot deal properly with SC electrons, due to the classical suppression of small momentum transfer collisions (cf. Ref. \cite{rb93} and Chapter 3 in \cite{sdr97}).

Since a major focus of this work is to look into the saturation behavior in high-energy collisions, the systems have been plotted with increasing projectile charge to velocity ratio $Z_p/v_p$, so that the systematic effects which would increase for high $Z_p/v_p$ can be studied in detail in the angular and energy distributions. 
\subsubsection{DDCS for net ionization as a function of electron emission angle at 200 eV\label{200ev}}
We focus on $E_{\rm{el}}=$ 200 eV for two reasons: first, the ionization mechanism producing high-energy electrons can be considered classical, and therefore CTMC should be working properly for this emission energy. Second, the shape of the DDCS at this emission energy is directly related to the influence of both target and projectile.

The usefulness of classical-trajectory calculations of ionization and capture has been established for simple atomic collision systems and for heavy-ion collisions, particularly when the collisions are deemed non-perturbative in nature and when quantum tunneling phenomena play a minor role. However, as demonstrated in \cite{rb93,kgh+95} the classical calculation of ionization is compromised for collisions in which the projectile undergoes small momentum transfer (less than about 0.7 a.u.). This weakness of the method is irrelevant for the emission energy of $E_{\rm{el}}=$ 200 eV, for which momentum transfer and ionized electron energy are strongly correlated in accord with the so-called Bethe ridge \cite{rb93}, but may have repercussions for the case of $E_{\rm{el}}=$ 5 eV discussed in the next subsection. For the case of $E_{\rm{el}}=$ 200 eV the distribution over momentum transfers to the projectile is peaking at 3.75 a.u., with a full width at half maximum (FWHM) of 2 a.u., as compared to a mean electron momentum of 3.8 a.u.

\begin{figure}
{\centerline{\includegraphics[width=0.7\linewidth]{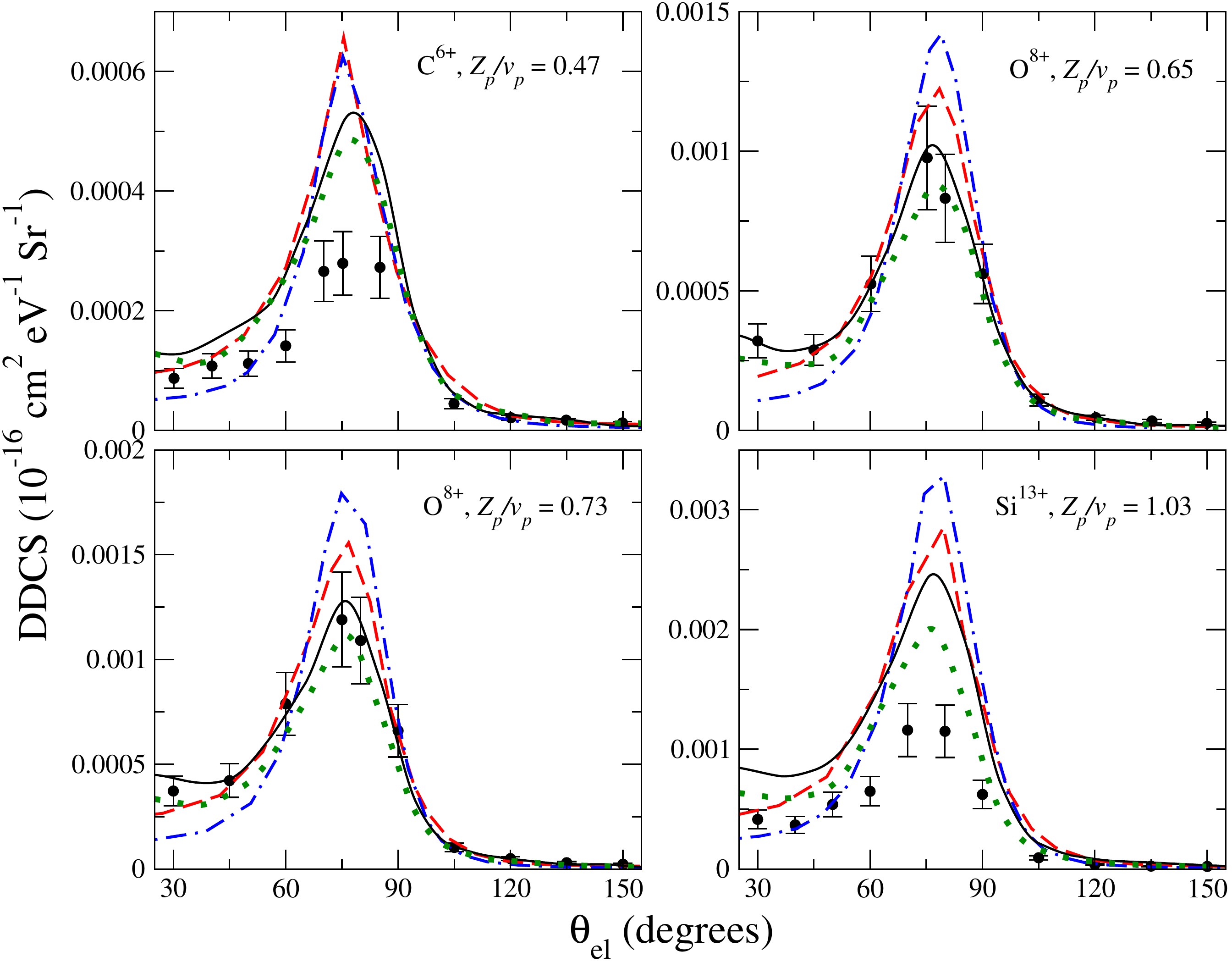}}}
\caption{DDCS for net ionization as a function of electron emission angle in degrees, for the electron emission energy $E_{\rm{el}}=$ 200 eV. Present CTMC results (obtained with angular resolution $\Delta \theta_{\rm{el}}=10^\circ$) are shown as full black and green dotted lines, with static and with time-dependent screening, respectively. The experimental data are shown as solid circles with error bars, and the prior and post CDW-EIS results are given by broken red lines and dash-dotted blue lines, respectively \cite{bbb+16,bbm+17,bbc+18}.\label{ddcs200}}
\end{figure}

In Fig. \ref{ddcs200} we display the experimental data and CDW-EIS calculations from \cite{bbb+16,bbm+17,bbc+18} with our CTMC results obtained with static and with time-dependent screening. We find an inconsistency in the comparison of CTMC data with measurements. For the O$^{8+}$ projectiles, the CTMC results present an almost perfect quantitative agreement with experiment, showing an accurate shape in forward/backward emission and in the maximum. In the case of C$^{6+}$ and Si$^{13+}$ projectiles the CTMC data show an overestimation with respect to the measured data. A closer inspection of these two systems shows that almost constant ratios prevail between CTMC and measured data for the whole range of emission angles.

When comparing with the perturbative CDW-EIS data, we find a general trend for the comparison of the theoretical data sets for the four systems. At forward angles the CTMC results are slightly higher than the perturbation theory results, while in the peak zone the classical results are lower, being approximately at 80\% and 70\% of the post and prior results, respectively. At backward angles no significant differences are found. In all cases, the perturbative results overestimate the measurements, but this overestimation is much greater in the case of the C$^{6+}$ and Si$^{13+}$ projectiles. Taking all this into account, one might attribute the differences for these two projectiles to a problem with the normalization factor in the measurements in \cite{bbm+17}. We will keep this in mind when looking at the emission energy 5 eV in subsection \ref{ddcs5sub}.  

Another important piece of information is contained in the forward-backward asymmetry in the experimental and theoretical results for $E_{\rm{el}}=$ 200 eV. There is very good agreement between CTMC data and the measurements for the oxygen projectiles. This quantitative agreement relies not only on the amount of ionized electrons, but also on the observed shape in the DDCS. This can also be extended to the C$^{6+}$ and Si$^{13+}$ projectiles, for which almost constant ratios are observed between CTMC and experimental data. It is known that the CDW-EIS method includes the effects coming from both the target and projectile, and therefore can be expected to work properly for the TCEE mechanism if a single-center description of the molecule in the exit channel suffices. CTMC also considers explicitly the effect of the two centers in the dynamics, but as implemented in this work it can also deal with the multi-center nature of H$_2$O. A close inspection shows that for these systems the CTMC results show a better reproduction of the experimental forward-backward asymmetry than the CDW-EIS data. We observe the ratio between the cross sections at 30$^{\circ}$ and 75$^{\circ}$ (80$^{\circ}$ in the case of Si$^{13+}$), which is independent of the absolute values, and thus independent of possible problems with the normalization of the experimental data. We show the values of the four systems in Table \ref{tablita}. They demonstrate that a CTMC calculation with an improved representation of the molecular potential leads to rather close values of this ratio to the experimentally observed value of about one third.

\begin{table}
\begin{center}
\begin{tabular}{@{}lccccc}\\
 \hline\noalign{\smallskip}
$Z_p/v_p$ (a.u.) &0.47 (C$^{6+}$)&0.65 (O$^{8+}$)&0.73 (O$^{8+}$)&1.03 (Si$^{13+}$)\\
\noalign{\smallskip}\hline\noalign{\smallskip}
Measurements&0.31&0.33&0.31&0.36\\
CTMC static screening&0.25&0.31&0.34&0.34\\
CTMC time-dependent screening&0.25&0.29&0.29&0.33\\
CDW-EIS prior&0.15&0.16&0.18&0.17\\
CDW-EIS post&0.089&0.083&0.086&0.084\\
\noalign{\smallskip}\hline
\end{tabular}
\end{center}
\caption{Ratios of the DDCS for 200 eV between forward (30$^{\circ}$) and peak ejection angles ($75^{\circ}$ except for Si$^{13+}$, where we use 80$^{\circ}$). Column 1: C$^{6+}$ at 4 MeV/u; 2: O$^{8+}$ at 3.75 MeV/u; 3: O$^{8+}$ at 3 MeV/u; 4: Si$^{13+}$ at 4 MeV/u. \label{tablita}}
\end{table}
The prior version of CDW-EIS is deemed to be better in dealing with the target center, since in the post version the term including the residual interaction between the passive and active electrons is not included \cite{bbb+16}.  However, in both versions the nuclear target potential experienced by the active continuum electron is represented by a {\it single} Coulomb potential with an effective charge. By looking at the results, it appears to us that not only a correct representation of the passive target electrons has to be considered, but also a proper description of the multi-center aspect of the molecule. That is, in order to obtain the correct forward-backward asymmetry in ion-molecule collisions a complete multi-center approach should be used. In our CTMC model we consider the interaction of electrons transferred to the continuum with an effective potential for the water molecule that incorporates a full three-center geometry, and, consequently, the forward-backward asymmetry is reproduced accurately. The TCCE mechanism, appropiate for ion-atom collisions, has to be extended to a MCEE mechanism when dealing with ion-molecule collisions. We observe that the two present model calculations (without and with time-dependent screening) yield similar results for the ratio between forward and peak ejection angle.
\subsubsection{DDCS for net ionization as a function of electron emission angle at 5 eV\label{ddcs5sub}}
We now consider the case of $E_{\rm{el}}=$ 5 eV for which the DDCS looks quite different to the previous one in that it has a relatively flat shape. The use of classical trajectories is not unproblematic in this case, since the underestimation of ionization in collisions with small momentum transfer to the projectile has been studied in the context of p-H and p-H$_2$ collisions \cite{kgh+95}. In the case of p-H collisions it is possible to add a perturbation theory contribution to account for these
large-impact parameter collision contributions. For the present heavy-ion system this is not straightforward, since the CDW-EIS calculations would have to be analyzed with respect
to momentum transfer. We looked at CTMC distributions over momentum transfer for ionizing events with electron emission at $E_{\rm{el}}=$ 5 eV. In contrast to the fast electron emission case
discussed in Sec. \ref{200ev}, we find the classical distribution to be skewed towards higher momentum transfers with a maximum at 0.85 a.u. and FWHM of 0.9 a.u., as compared to an average ionized electron momentum value of 0.61 a.u.. This reflects a lack of ionization events producing $E_{\rm{el}}=$ 5 eV electrons in large-impact-parameter collisions. We estimate the effect to be responsible for a $\sim$20\% reduction in the DDCS with a bigger effect at backward angles.  

Given this shortfall the apparent agreement between CTMC and experiment for both C$^{6+}$ and Si$^{13+}$ therefore should be considered fortuitous. This reinforces the idea that there is a potential problem with the normalization of the (experimental) data from \cite{bbm+17} for C$^{6+}$ and Si$^{13+}$ projectiles. This problem can also be seen in the comparison between CDW-EIS results and measurements, which shows very good agreement for O$^{8+}$ projectiles but not for C$^{6+}$ and Si$^{13+}$. Soft-collision electrons should be described properly with CDW-EIS and, as can be seen, the measurements for O$^{8+}$ projectiles lie in general between the post and prior versions. It is worth noting that a direct comparison between all theoretical sets shows the expected general behavior for the four systems, with an underestimation of the classical results for this small electron energy with respect to the perturbative results. Nevertheless, we have to keep in mind that for highly charged ions the treatment of screened electrons by a static potential may be more problematic and therefore a further analysis needs to be performed. This is the motivation for focusing on the time-dependent screening effect in the following sections.
\begin{figure}
{\centerline{\includegraphics[width=0.7\linewidth]{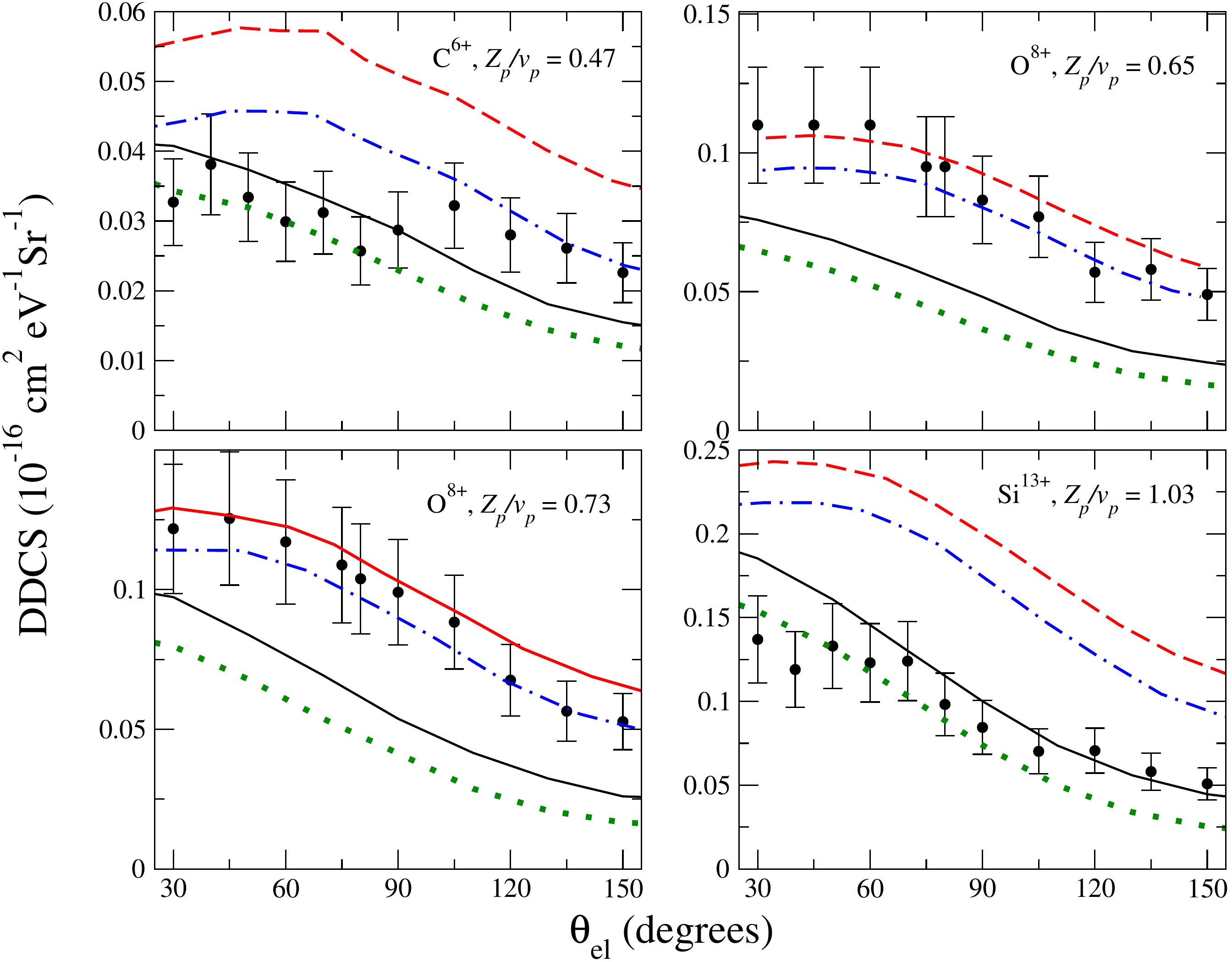}}}
\caption{Same as Fig. \ref{ddcs200} but for the electron emission energy $E_{\rm{el}}=$ 5 eV. The angular resolution for the present CTMC calculations is chosen at $\Delta \theta_{\rm{el}}=20^\circ$. \label{ddcs5}}
\end{figure}
\subsubsection{Time-dependent screening in the DDCS for net ionization as a function of electron emission angle}
An important fact in fast ion-atom and ion-molecule collisions is that an increase in the projectile charge leads to a significant decrease in the electron emission at backward angles (cf. Chapter 2 in \cite{sdr97}) due to the attraction of the emitted electrons by the projectile. We find that the time-dependent screening introduced in our model affects predominantly the emission of electrons at backward angles. In the case of $E_{\rm{el}}=$ 5 eV (Fig. \ref{ddcs5}), the screening reduces the emission in the forward direction by about 15\% for all the systems, while the reduction at backward angles is not only stronger but also depends on the projectile charge and velocity: the reduction is by about 20\%, 30\%, 35\% and 45\% for the backward angles for C$^{6+}$ ($Z_p/v_p=0.47$), O$^{8+}$ ($Z_p/v_p=0.65$), O$^{8+}$ ($Z_p/v_p=0.73$) and Si$^{13+}$ ($Z_p/v_p=1.03$), respectively. Thus, we find that in the forward direction the screening only affects the amount of ionization produced, while in the backward direction the target center influence is important and a correct description of its ionization state is necessary. In general, the relative difference between results with static and with time-dependent screening increases smoothly with increasing emission angle.

For high-emission-energy electrons, where the ionization can be considered classical, more conclusions can be obtained. If we focus on the DDCS at $E_{\rm{el}}=$ 200 eV, for the O$^{8+}$ projectiles we find good agreement of both CTMC variants with respect to measurements. This agreement gets better with the static screening in the case of O$^{8+}$ ($Z_p/v_p=0.65$), while for O$^{8+}$ ($Z_p/v_p=0.73$) the measurements lie in the middle of the two classical sets (static vs time-dependent), as shown in Fig. \ref{ddcs200}. Therefore, a better understanding of the saturation effect at high $Z_p/v_p$ would require more experimental data for $Z_p/v_p\gtrsim 1$. 
\subsubsection{DDCS for net ionization as a function of electron emission angle separated in the five MO contributions}
\begin{figure}
{\centerline{\includegraphics[width=0.7\linewidth]{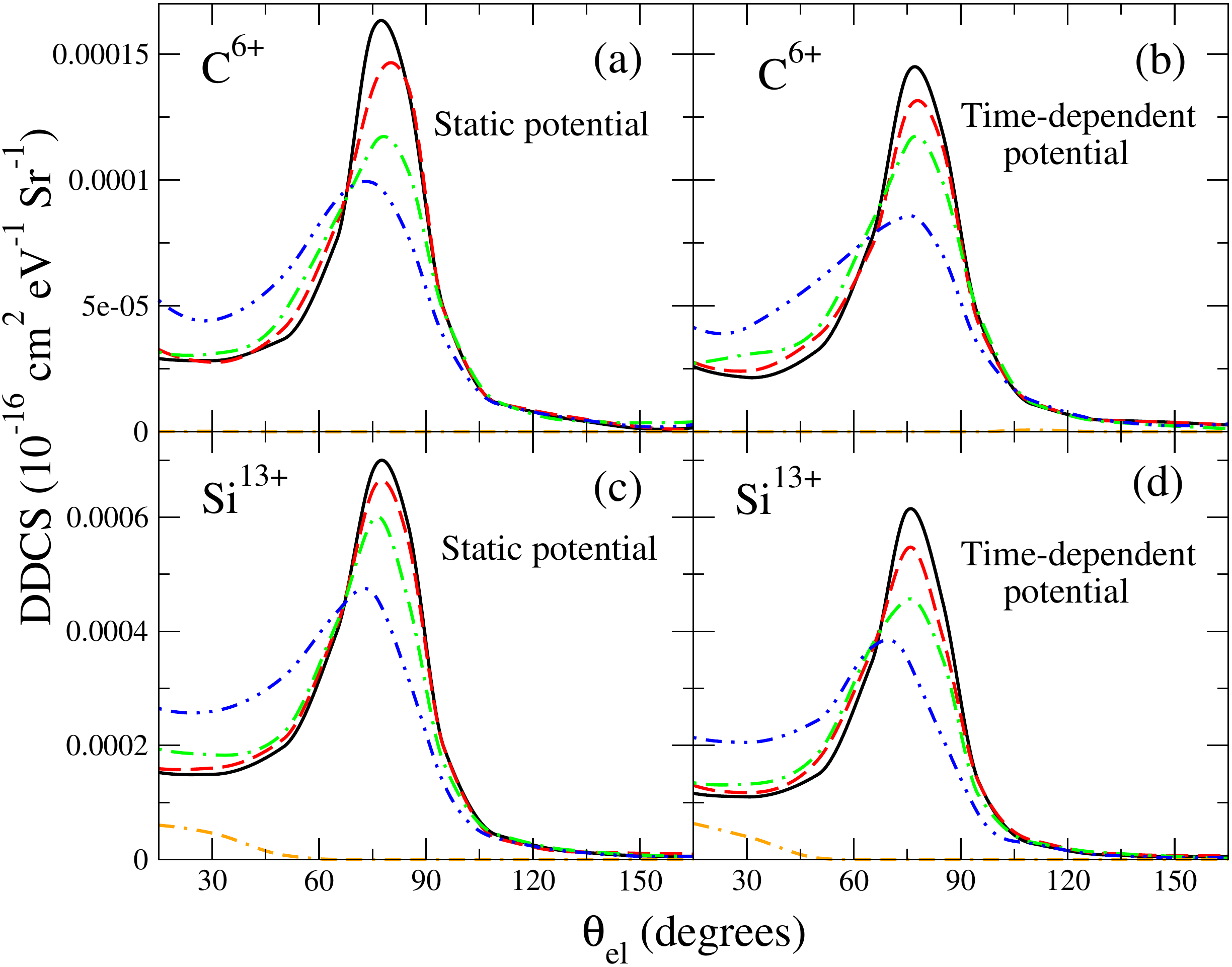}}}
\caption{DDCS for net ionization as a function of the electron emission angle in degrees for $E_{\rm{el}}=$ 200 eV separated into the five MO contributions. In the upper panels results are shown for C$^{6+}$ and in the lower ones for Si$^{13+}$ projectiles, both with initial velocity of 12.65 a.u.; with static (a), (c) and with time-dependent (b), (d) screening. (---) $1b_1$, ({\textcolor{red}{$--$}}) $3a_1$, ({\textcolor{green}{$\cdot-\cdot$}}) $1b_2$, ({\textcolor{red}{$\cdot\cdot-\cdot\cdot$}}) $2a_1$, ({\textcolor{orange}{$--\cdot--$}}) $1a_1$.\label{ddcs20mo6-13}}
\end{figure}

It is also interesting to look at the contributions from each MO for a given differential cross section. We focus again on the DDCS as a function of the emission angle for $E_{\rm{el}}=$ 200 eV. In Fig. \ref{ddcs20mo6-13} we plot the contributions from each MO for the C$^{6+}$ and Si$^{13+}$ projectiles with static and with time-dependent screening. We find a similar behavior for the two systems considered: (i) the main contribution from the $2a_1$ MO is in forward emission, (ii) in the peak the contribution from each MO depends on the ionization potential, and (iii) similar contributions (except for the $1a_1$) are found at backward angles. 

The $1a_1$ is only of importance at forward emission in the case of Si$^{13+}$, since its ionization potential is one order of magnitude larger than for the rest of MOs. An increasing contribution for higher emission energies can be expected from $1a_1$, as stated in \cite{tmf+14}, due to the higher electron speed and the fact that the contribution comes from small impact parameters, i.e., from direct penetration of the projectile in this inner region. 

The reduction due to the time-dependent screening approach tends to be similar for all the MOs: in the case of C$^{6+}$ around 15\% and between 24-35\% for the Si$^{13+}$ projectile. Therefore, the time-dependent screening affects roughly all the orbitals in the same way.
\subsubsection{SDCS for net ionization as a function of electron emission angle}
\begin{figure}
{\centerline{\includegraphics[width=0.55\linewidth]{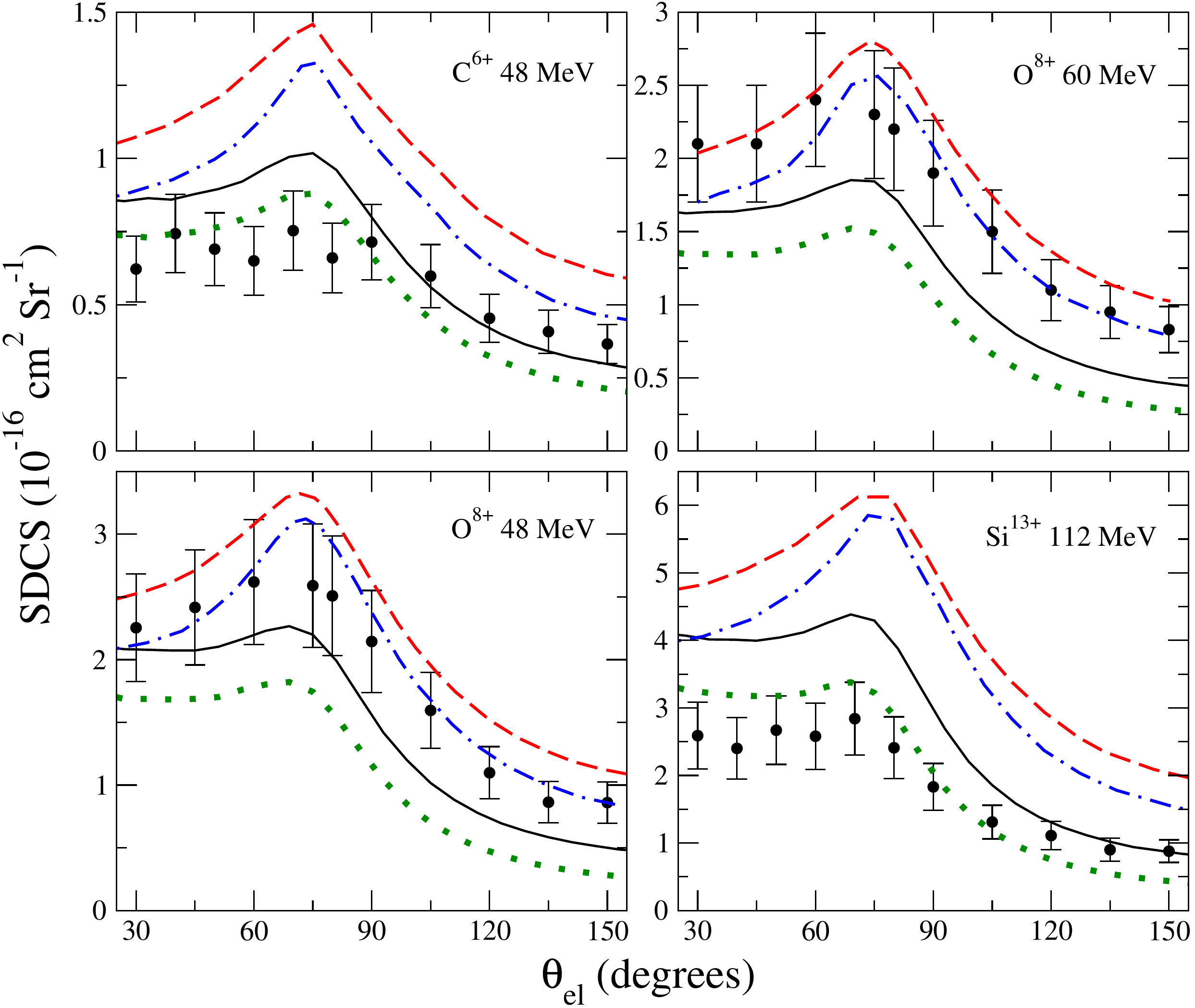}}}
\caption{SDCS for net ionization as a function of electron emission angle in degrees. Present CTMC results are shown as full black and green dotted lines, with static and with time-dependent screening, respectively. The experimental data are shown as solid circles with error bars, and the prior and post CDW-EIS results are given by broken red lines and dash-dotted blue lines, respectively \cite{bbb+16,bbm+17,bbc+18}. The angular resolution for the present CTMC calculations is chosen at $\Delta \theta_{\rm{el}}=6^\circ$.\label{sdcstheta}}
\end{figure}

In Fig. \ref{sdcstheta} the SDCS as a function of the ejection angle are shown. We find again that if we look at the ratio between the cross section at 30$^{\circ}$ and 75$^{\circ}$ for the different sets of data, as in table \ref{tablita}, the CTMC values agree better with the measurements than the CDW-EIS calculations. Within these, the prior version ratios are better than the post ones. We find again that the more accurate the description of the target field is, the better the forward-backward assymetry. We think that a proper representation of the multi-center aspect of the water molecule does have an important effect, as stated before, namely that the mechanism is MCEE as opposed to TCEE.

In terms of a quantitative comparison, we expect a general underestimation for both (static and time-dependent screening) CTMC models in the whole angular range, since low-energy electrons, produced in soft collisions, are emitted over a wide range of angles. Therefore, the expected behavior of the CTMC SDCS as a function of the emission angle is showing the correct shape, but the height is underestimated. This is the behavior found for the O$^{8+}$ projectiles, but for the two other projectiles we find the opposite in terms of height. Concerning the saturation versus $Z_p/v_p$ discussed in Ref. \cite{bbc+18} (in the sense of a turn-over in the total cross section) we can state that neither the CDW-EIS nor our present calculations support such a behavior.

The results for the two CTMC methods  shown in Fig. \ref{sdcstheta}, namely with static screening (black line) and time-dependent screening (dotted green line) show how there is a  gradual increase in the effect from the improved screening model as $Z_p/v_p$ increases. While this model is apparently in good agreement for the smallest and largest values of this parameter (while the CDW-EIS results reach much higher values in these cases) the situation is very different for the oxygen projectiles and the intermediate values of $Z_p/v_p$ (for which CDW-EIS agrees, and the CTMC results show smaller SDCS values). This situation warrants further investigation, since the $Z_p/v_p$ parameter is considered to be a useful parameter to characterize the high-energy collisions ionization dynamics.
\subsubsection{SDCS for net ionization as a function of electron emission angle separated into the first five $q-$fold contributions}
\begin{figure}
{\centerline{\includegraphics[width=0.7\linewidth]{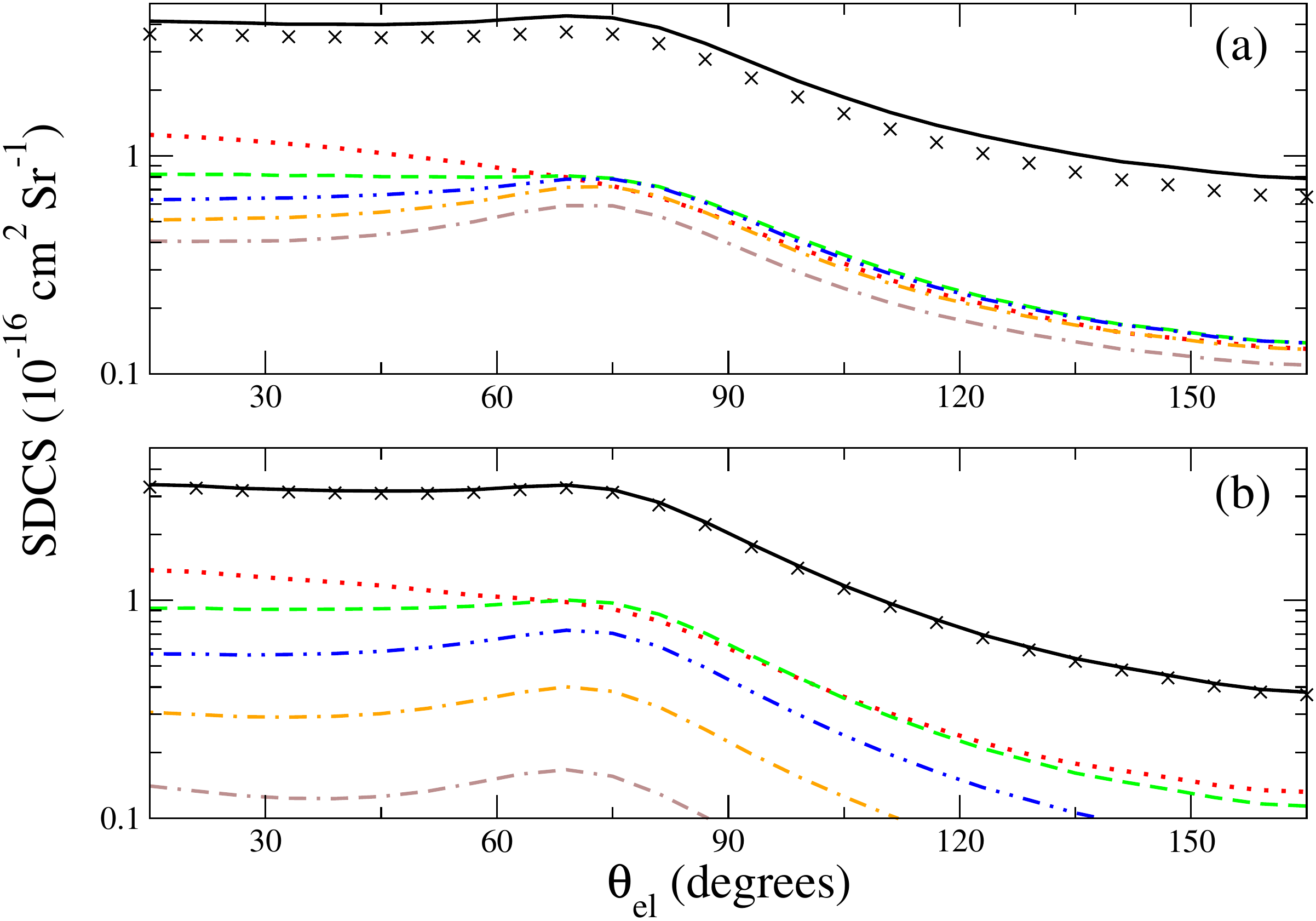}}}
\caption{SDCS for net ionization as a function of electron emission angle in degrees for Si$^{13+}$, with static (a) and with (b) time-dependent screening at 112 MeV. (---) Net, ({\textcolor{red}{$\cdot\cdot\cdot$}}) $\frac{{\rm{d}} \sigma_1}{{\rm{d}} \Omega_{\rm{el}}}$, ({\textcolor{green}{$--$}}) $2\frac{{\rm{d}} \sigma_2}{{\rm{d}} \Omega_{\rm{el}}}$, ({\textcolor{red}{$\cdot\cdot-\cdot\cdot$}}) $3\frac{{\rm{d}} \sigma_3}{{\rm{d}} \Omega_{\rm{el}}}$, ({\textcolor{orange}{$\cdot-\cdot$}}) $4\frac{{\rm{d}} \sigma_4}{{\rm{d}} \Omega_{\rm{el}}}$, ({\textcolor{brown}{$--\cdot--$}}) $5\frac{{\rm{d}} \sigma_5}{{\rm{d}} \Omega_{\rm{el}}}$, ($\times$) $\sum\limits_{q=1}^5q\frac{{\rm{d}} \sigma_q}{{\rm{d}} \Omega_{\rm{el}}}$.\label{sdcstq}}
\end{figure}

We now look at the contribution from the $q-$fold terms to the SDCS, $\frac{{\rm{d}} \sigma_{\rm{Net}}}{{\rm{d}} \Omega_{\rm{el}}}=\sum\limits_{q=1}^{10}q\frac{{\rm{d}}\sigma_q}{{\rm{d}} \Omega_{\rm{el}}}$. In Fig. \ref{sdcstq} we show the contributions from the first five $q\frac{{\rm{d}} \sigma_q}{{\rm{d}} \Omega_{\rm{el}}}$ terms, as well as the sum truncated at $q=5$ (shown with crosses), compared to the net cross section (full line), for the Si$^{13+}$ system. Within the time-dependent screening method this sum of the first five terms is sufficient to reach around 97\% of the net cross section for the whole angular range. In contrast, with static screening this contribution goes from 87\% at forward emission to 82\% at backward angles. 
By looking at the individual contributions, we find that only slight differences are found for the $\sigma_1$ and $2\sigma_2$ terms. The higher terms are expectedly the most affected by the time-dependent screening approach, which shows that this method suppresses high-order multiple ionization. With respect to the angular dependence, time-dependent screening $q >$ 3 terms are smaller than their static counterparts in the entire range, with greater differences in the backward direction. In order to test further the construction of the time-dependent screening approach it would be of great interest to have access to experimental data of differential electron emission in coincidence with the charge state produced during the collision.
\subsection{Energy distributions}
\subsubsection{DDCS for net ionization as a function of electron emission energy}
\begin{figure}
{\centerline{\includegraphics[width=0.6\linewidth]{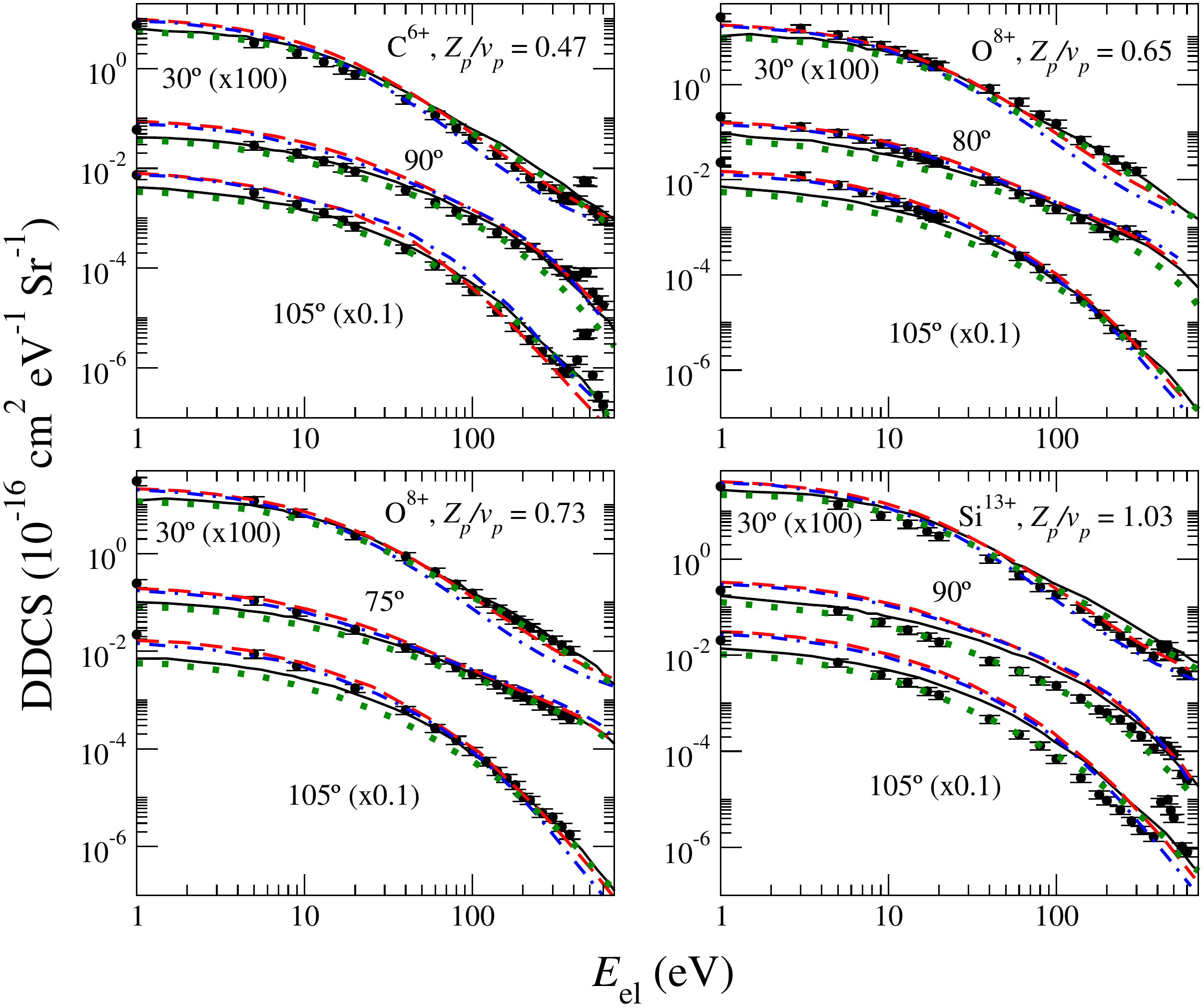}}}
\caption{DDCS for net ionization  as a function of the electron energy. Present CTMC results are shown as full black and green dotted lines, with static and with time-dependent screening, respectively. The experimental data are shown as solid circles with error bars, and the prior and post CDW-EIS results are given by broken red lines and dash-dotted blue lines, respectively. The energy resolution for the CTMC calculation is $\Delta E_{\rm{el}}=1$ eV for $E_{\rm{el}}<10$ eV; $\Delta E_{\rm{el}}=10$ eV for $10<E_{\rm{el}}<100$ eV and $\Delta E_{\rm{el}}=100$ eV for $100<E_{\rm{el}}<1000$ eV. \cite{bbb+16,bbm+17,bbc+18}.\label{ddcse}}
\end{figure}

In Fig. \ref{ddcse} the DDCS are shown as a function of the electron energy $E_{\rm{el}}$ for three representative ejection angles. A direct comparison between CDW-EIS and CTMC data shows some general trends for all four systems. At low $E_{\rm{el}}$ there is an underestimation of CTMC data with respect to CDW-EIS results, which becomes larger with increasing emission angle. 

In the intermediate- and high-energy range we find in general agreement between all the theory sets, except for 30$^\circ$ at the highest energies shown, where the CTMC results are higher than the CDW-EIS values. These high-energy electrons emitted in the forward direction are likely to be better described with a multi-center potential for the target, since electrons moving in this direction are the most affected by the target potential.
The comparisons with measurements in these angular and energy ranges show very good agreement with CTMC for the O$^{8+}$ projectiles, while very good agreement is found between CDW-EIS and experiments for C$^{6+}$ and Si$^{13+}$ projectiles. This is analogous to what was found for the cross sections as a function of emission angle.
\subsubsection{Time-dependent screening in the DDCS for net ionization as a function of electron emission energy}
Comparing the two screening models (static vs time-dependent) two main observations can be made. First, we find that the relative differences are greater in the intermediate emission energy (10-100 eV) than in the low- and high-energy ranges. Second, the relative differences are smaller at 30$^{\circ}$ in the whole energy range than for 75$^{\circ}$ and 105$^{\circ}$. Therefore, taking into account that for backward emission and intermediate-energy emission the many-center mechanism is mainly responsible for the detailed shape of the cross sections, we find that the time-dependent screening affects the MCEE mechanism the most. The small reduction for low-energy electrons in all considered systems means that soft electrons are not affected significantly, since they come mainly from single ionization (as can be seen in Fig. \ref{sdcseq}, where the SDCS is shown for the Si$^{13+}$ projectile, divided in $q$-fold terms up to $q=5$).  
\subsubsection{SDCS for net ionization as a function of electron emission energy}
\begin{figure}
{\centerline{\includegraphics[width=0.6\linewidth]{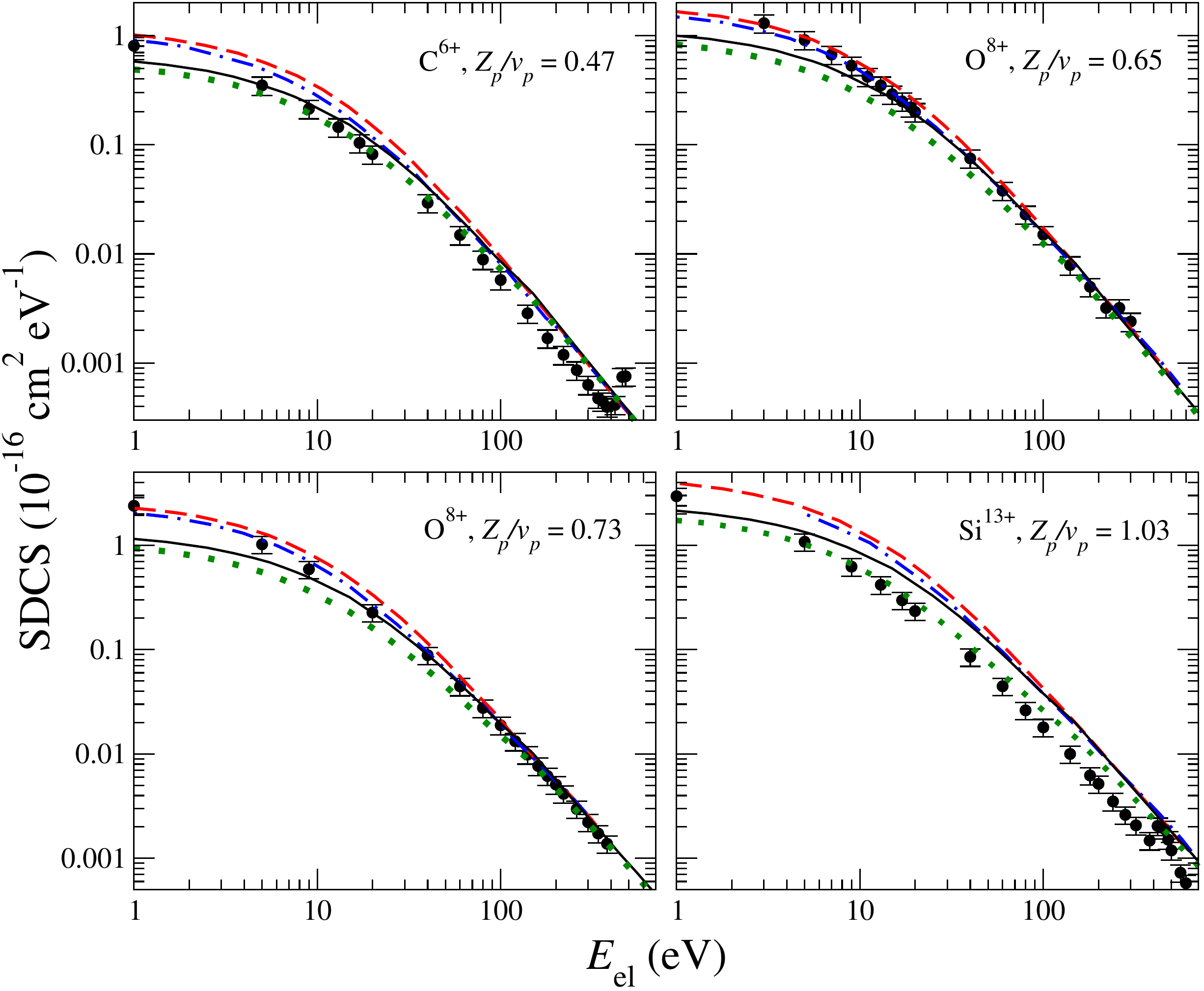}}}
\caption{SDCS for net ionization as a function of the electron energy. Present CTMC results are shown as full black and green dotted lines, with static and with time-dependent screening, respectively. The experimental data are shown as solid circles with error bars, and the prior and post CDW-EIS results are given by broken red lines and dash-dotted blue lines, respectively \cite{bbb+16,bbm+17,bbc+18}. The energy resolution for the CTMC calculation is $\Delta E_{\rm{el}}=1$ eV for $E_{\rm{el}}<10$ eV; $\Delta E_{\rm{el}}=10$ eV for $10<E_{\rm{el}}<100$ eV and $\Delta E_{\rm{el}}=100$ eV for $100<E_{\rm{el}}<1000$ eV.\label{sdcse}}
\end{figure}

The agreement between CTMC data and O$^{8+}$ experiments for the DDCS at high-emission energy can also be observed in the SDCS plot in Fig. \ref{sdcse}. There is also perfect agreement between CDW-EIS and CTMC results in this energy range ($E_{\rm{el}}>100$ eV) for all four systems. The agreement found with respect to the experiments for the O$^{8+}$ projectiles and the overestimation for C$^{6+}$ and Si$^{13+}$ systems again suggest that there is a normalization problem with the experiments for the two latter systems. Since this difference is much more pronounced for the Si$^{13+}$ projectile than for C$^{6+}$, and looking at the reduction found with the time-dependent screening results, it is clear that for $Z_p/v_p\sim 1$ there is strong evidence for non-perturbative behavior that is captured by the time-dependent screening model in our calculations.
\subsubsection{SDCS for net ionization as a function of electron emission energy separated in the first five $q-$fold contributions}
\begin{figure}
{\centerline{\includegraphics[width=0.7\linewidth]{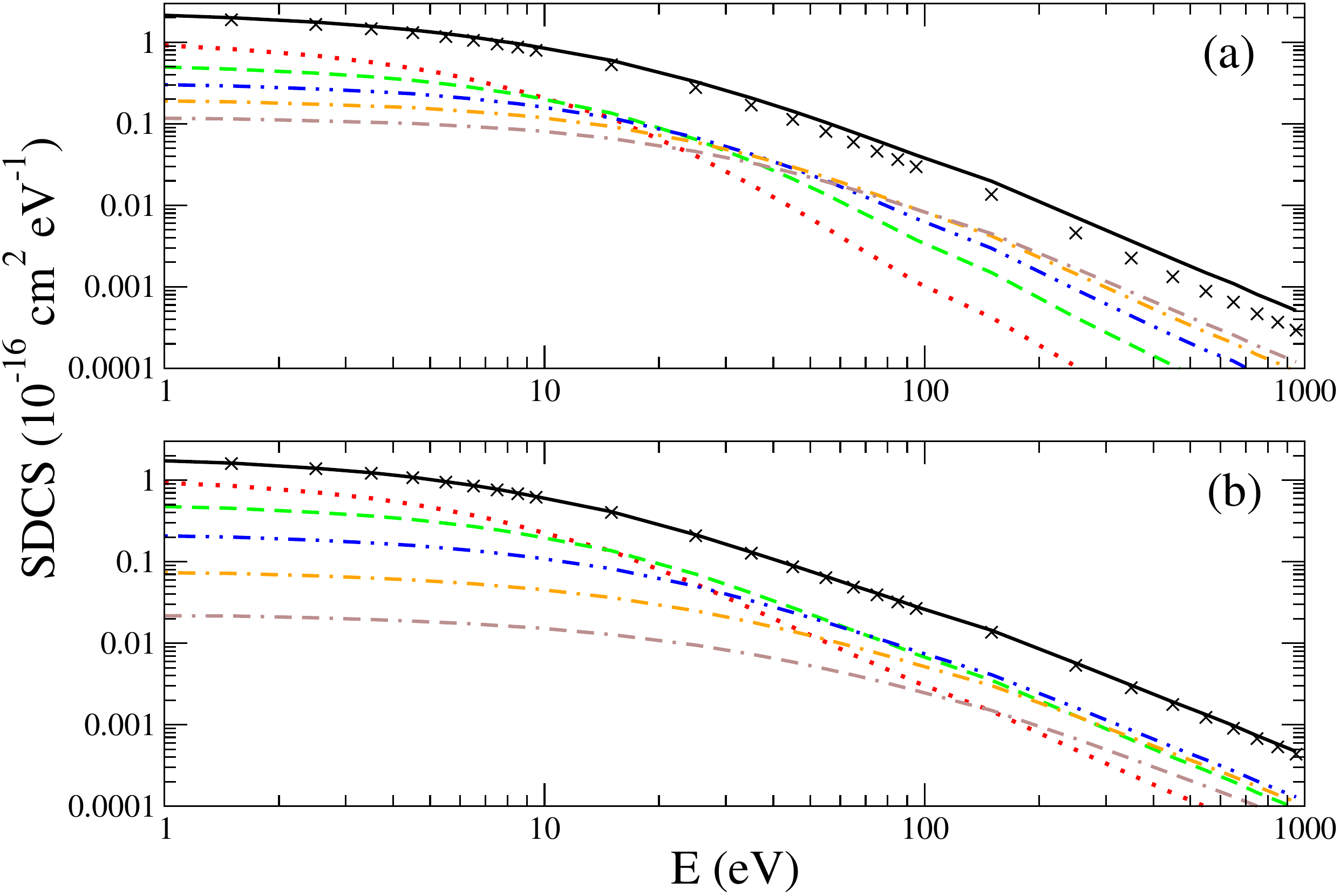}}}
\caption{SDCS for net ionization as a function of electron energy for Si$^{13+}$, with static (a) and with time-dependent (b) screening at 112 MeV. (---) Net, ({\textcolor{red}{$\cdot\cdot\cdot$}}) $\frac{{\rm{d}} \sigma_1}{{\rm{d}} E}$, ({\textcolor{green}{$--$}}) $2\frac{{\rm{d}} \sigma_2}{{\rm{d}} E}$, ({\textcolor{red}{$\cdot\cdot-\cdot\cdot$}}) $3\frac{{\rm{d}} \sigma_3}{{\rm{d}} E}$, ({\textcolor{orange}{$\cdot-\cdot$}}) $4\frac{{\rm{d}} \sigma_4}{{\rm{d}} E}$, ({\textcolor{brown}{$--\cdot--$}}) $5\frac{{\rm{d}} \sigma_5}{{\rm{d}} E}$, ($\times$) $\sum\limits_{q=1}^5q\frac{{\rm{d}} \sigma_q}{{\rm{d}} E}$.\label{sdcseq}}
\end{figure}

We examine the importance of the first five $q-$fold contributions to the SDCS in Fig. \ref{sdcseq} for both CTMC approaches for the highest $Z_p/v_p$ studied system. In the low-energy region the contribution from terms up to $q=5$ is greater than 90\% in both cases. When the time-dependent screening approach is applied, this percentage stays constant over the whole energy range, while with a static target potential it decreases to 55\% for the highest energies shown. Therefore, with static screening, the contribution from the $q>5$ terms is almost dominant for high-energy electrons, which is not compatible with an IEM description based on an H$_2$O$^+$ potential. The relative differences of the two net cross sections are around 20\% and 30\% for the lowest and highest emission energies, respectively, but these differences come from different mechanisms. A gradual decrease of the $q-$fold terms with increasing $q$ is found for the low-energy range, while in the high-energy range it only happens for $q>4$, whereas the lower-$q$ terms are stronger than their frozen static potential counterparts. This increase is especially striking in the case of the single ionization term, whose contribution at high energy is increased by more than a 100\% for $E_{\rm{el}}>100$ eV when time-dependent screening is applied. 
\section{Conclusions\label{conclusiones}}
We have studied the validity of the CTMC model as applied to the calculation of SDCS and DDCS for net ionization in fast highly-charged ion collisions with water molecules by comparing results to recent experimental and theoretical data. We also note that the present CTMC calculations contain an average over all molecular orientations. We observe that the classical description compares well with the available data except for the lowest electron emission energies analyzed. We have also investigated the saturation behavior (deviation of the measured ionization from the Bethe-Born model prediction) apparent in the experimental data for high $Z_p/v_p$, in terms of both singly and doubly differential cross sections. To this end, we have implemented a time-dependent screening approach that takes into account the removal of electrons during the collision to update the charge state of the target at each time step, in the spirit of time-dependent mean field theory. 

We conclude that the main decrease in ionization due to the time-dependent screening happens for backward emission and for intermediate energies in the differential cross sections as a function of the emission angle and energy, respectively. Therefore the two/many-center electron emission mechanism is the most affected by this improved treatment. 

We also observe that the effect of the time-dependent screening becomes more important for increasing $Z_p/v_p$, and that it is relevant in the 112 MeV Si$^{13+}$ system. Nevertheless, more experimental work is required to establish a systematic dependence of the differential cross sections as a function of $Z_p/v_p$ in order to shed more light on the deviation from Bethe-Born type scaling.

The angular distribution for intermediate to high energies can be estimated by CTMC properly, which reproduces the asymmetry in backward to forward directions quite well. It also describes the peak maximum zone and shows for the highest energies better agreement with the experimental values than the CDW-EIS results. This suggests that a three-center potential description is required when dealing with ion-water molecule collisions for the angular distribution of the emitted electrons. We have named the process MCEE due to the effect coming from the multiple atomic centers in the target molecule in addition to the projectile. More highly differential cross sections than the DDCS presented here might reveal bigger differences.
\section{Acknowledgments}
This work was supported by the York Science Fellowship program, the Natural Sciences and Engineering Research Council of Canada (NSERC) and by Ministerio de Econom\'ia y Competitividad (Spain) (Project No. FIS2017-84684-R). High-performance computing resources for this work were provided by Compute Canada/Calcul Canada.

\end{document}